\documentclass[copyright,creativecommons]{eptcs}
\providecommand{\event}{PDMC 2009}
\providecommand{\volume}{??}
\providecommand{\anno}{2009}
\providecommand{\firstpage}{1}

\usepackage{amssymb}
\usepackage{colortbl}
\usepackage{amsmath}
\usepackage{amsxtra}
\usepackage{amstext}
\usepackage{latexsym}
\usepackage{setspace}
\usepackage[dvips]{graphicx}
\usepackage{breakurl}
\usepackage{color}

\begin{document}

\title { An Efficient Explicit-time Description Method for Timed Model Checking}
\author{ Hao Wang and Wendy MacCaull
         \institute{Centre for Logic and Information\\
         St. Francis Xavier University\\
         Antigonish, Canada }
         \email{\{hwang, wmaccaul\}@stfx.ca}
}

\def\titlerunning{An Efficient Explicit-time Description Method for Timed Model Checking}
\def\authorrunning{H. Wang \& W. MacCaull}

\maketitle

\begin{abstract}
\emph{Timed} model checking, the method to formally verify real-time systems, is attracting increasing attention from both the model checking community and the real-time community. \emph{Explicit-time description methods} verify real-time systems using general model constructs found in standard un-timed model checkers. Lamport proposed an explicit-time description method \cite{Lamport05TRrealSimple} using a clock-ticking process (\emph{Tick}) to simulate the passage of time together with a group of global variables to model time requirements. Two methods, the \emph{Sync-based Explicit-time Description Method} using rendezvous synchronization steps and the \emph{Semaphore-based Explicit-time Description Method} using only one global variable were proposed \cite{Hao09SEDM,Hao09SMEDM}; they both achieve better \emph{modularity} than Lamport's method in modeling the real-time systems. In contrast to timed automata based model checkers like UPPAAL \cite{bllpwDimacs95uppaal}, explicit-time description methods can access and store the current time instant for future calculations necessary for many real-time systems, especially those with pre-emptive scheduling. However, the \emph{Tick} process in the above three methods increments the time by one unit in each tick; the state spaces therefore grow relatively fast as the time parameters increase, a problem when the system's time period is relatively long. In this paper, we propose a more efficient method which enables the \emph{Tick} process to leap multiple time units in one tick. Preliminary experimental results in a high performance computing environment show that this new method significantly reduces the state space and improves both the time and memory efficiency.
\end{abstract}

\section{Introduction}\label{SEC:introduction}

Model checking is an automatic analysis method which explores all possible states of a modeled system to verify whether the system satisfies a formally specified property. It was popularized in industrial applications, e.g., for computer hardware and software, and has great potential for modeling complex and distributed business processes. \emph{Timed} model checking, the method to formally verify real-time systems, is attracting increasing attention from both the model checking community and the real-time community. However, standard model checkers like SPIN \cite{Holzmann91BKspin} and SMV \cite{McMillan92THsmv} can generally only represent and verify the \emph{qualitative} relations between events, which constrains their use for real-time systems. \emph{Quantified} time notions, including time instant and duration, must be taken into account for timed model checking. For example in a safety critical application such as in an emergency department, after an emergency case arrives at the hospital, standard model checking can only verify whether ``the patient receives a certain treatment'', but to save the patient's life, it should be verified whether ``the patient receives a certain treatment within 1 hour''.

Many formalisms with time extensions have been presented as the basis for timed model checkers. Two popular ones are: (1) \emph{timed automata} \cite{DBLP:journals/tcs/AlurD94}, which is an extension of finite-state automata with a set of clock variables to keep track of time; (2) \emph{time Petri Nets} \cite{Merlin74TCPNthesis}, which is an extension of the Petri Nets with timing constraints on the firings of transitions. Various translation methods have been presented between time Petri Nets to timed automata \cite{DBLP:conf/apn/PenczekP04TCPNtoTA} in order to apply time-automata-based methods to time Petri Nets. UPPAAL \cite{bllpwDimacs95uppaal} and KRONOS \cite{DBLP:journals/sttt/Yovine97kronos} are two well-known timed automata based model checkers; they have been successfully applied to various real-time controllers and communication protocols. Conventional temporal logics like \emph{Linear Temporal Logic} (LTL) or \emph{Computation Tree Logic} (CTL) must be extended \cite{DBLP:conf/rex/AlurH91} to handle the specification of properties of timed automata. In order to handle continuous-time semantics, specialized data structures are needed to represent real clock variables, e.g. Difference Bounded Matrices \cite{DBLP:conf/avmfss/Dill89DiffBndMatrix} (employed by UPPAAL and KRONOS).

The foundation for the decidability results in timed automata is based on the notion of \emph{region equivalence} over the clock assignment \cite{BengtssonY03timedAutomata}. Models in a timed automata based model checker can not represent at which time instant a transition is executed within a time region; such model checkers can only deal with a specification involving a time region or a pre-specified time instant and cannot store the exact time instant when the transition is executed. However, many real-time systems, especially those with pre-emptive scheduling, need this information for succeeding calculations. For example, triage is widely practiced in medical procedures; the caregiver \emph{C} may be administering some required but non-critical treatment on patient \emph{A} when another patient \emph{B} presents with a critical situation, such as a cardiac arrest. \emph{C} then must move to the higher priority task of treating \emph{B}, but it is necessary to store the elapsed time of \emph{A}'s treatment to determine how much time is still needed or else the treatment must be restarted. The \emph{stop-watch} automata \cite{DBLP:conf/tacas/AbdeddaimM02StopWatchA}, an extension of timed automata, is proposed to tackle this; unfortunately as Krc{\'a}l and Yi discussed in \cite{DBLP:conf/tacas/KrcalY04decidableTA}, since the reachability problem for this class of automata is undecidable, there is no guarantee for termination in the general case.

Lamport \cite{Lamport05TRrealSimple} advocated \emph{explicit-time description methods} using general model constructs, e.g., global integer variables or synchronization between processes commonly found in standard un-timed model checkers, to realize timed model checking. He presented an explicit-time description method, which we refer to as LEDM, using a clock-ticking process (\emph{Tick}) to simulate the passage of time, and a pair of global variables to store the time lower and upper bounds for each modeled system process. The method has been implemented with popular model checkers SPIN (sequential) and SMV. We presented two methods, (1) the \emph{Sync-based Explicit-time Description Method} (SEDM) \cite{Hao09SEDM} using rendezvous synchronization steps between the \emph{Tick} and each of the system processes; and (2) the \emph{Semaphore-based Explicit-time Description Method} (SMEDM) \cite{Hao09SMEDM} using only one global semaphore variable. Both these methods enable the time lower and upper bounds to be defined locally in system processes so that they provide better \emph{modularity} in system modeling and facilitate the use of more complex timing constraints. Our experiments \cite{Hao09SMEDM,Hao09SEDM} showed that the time and memory efficiencies of these two methods are comparable to that of LEDM.

The explicit-time description methods have three advantages over timed-automata-based model checkers: (1) they do \emph{not} need specialized languages or tools for time description so they can be applied in standard un-timed model checkers. Recently, Van den Berg et al. \cite{DBLP:conf/fmics/BergSW07LEDMcaseStudy} successfully applied LEDM to verify the safety of railway interlockings for one of Australia's largest railway companies; (2) they enable the accessing and storing of the current time \cite{Hao09SMEDM}, a useful feature for pre-emptive scheduling problems; and (3) they enable the usage of large-scale distributed model checkers, e.g., {\sc DiVinE}, for timed model checking.

Orthogonally, model checking has been studied in parallel and distributed computing platforms. Because real world models often come with gigantic state spaces which can not fit into the memory of a standard computer, inevitably a portion of the state space needs to be accessed from the secondary storage and the model checking algorithm becomes very slow \cite{Brim2004ErcimNews}. This problem is known as \emph{state explosion}. Large-scale analysis is needed in many practical cases. Distributed model checkers exploit the power of distributed computing facilities so that much larger memory is available to accommodate the state space of the system model; parallel processing of the states can, moreover, reduce the verification time. Our experiments \cite{Hao09SEDM} compared the time efficiency between the sequential SPIN and {\sc DiVinE} \cite{divinePrjPage}, a well-known distributed model checker. When using the same explicit-time description method, {\sc DiVinE} can verify much larger models and finish the verification for models of the same size in significantly less time than SPIN.

In this paper, we present a new explicit-time description method called \emph{Efficient Explicit-time Description Method} (EEDM). We found that the former three methods (LEDM, SEDM and SMEDM) suffer from one common problem: as the \emph{Tick} process increments the time by one unit in each tick, the state space grows relatively fast as the time parameters increase. E.g., in our experiment \cite{Hao09SMEDM} using LEDM, the number of states doubles as time bounds grow from 12 to 14. In the new EEDM, the \emph{Tick} can increment the time in two modes: the \emph{standard} mode and the \emph{leaping} mode. When it is necessary to store the current time to allow access for future calculations, it ticks in the standard mode; otherwise, it ticks in the leaping mode. For each system process, we define one global variable indicating whether the process needs to store and access the current time, allowing the \emph{Tick} process to switch between the standard mode and the leaping mode. For the experiments, we continue using {\sc DiVinE} (the method is also applicable to other standard model checkers); the results show that: in the leaping mode, the number of states can be reduced significantly, so it is much less affected by the increase of time parameters; in the standard mode, the time and memory efficiencies are comparable with the former methods.

The remainder of the paper is organized as follows. Section \ref{SEC:preliminary} gives background information with respect to the {\sc DiVinE} model checker. The new explicit-time description method implemented in {\sc DiVinE} is presented in Section \ref{SEC:timeDesc}; for comparison, LEDM is also briefly described in the same section. Section \ref{SEC:exprDesc} describes our experiments and the results. Section \ref{SEC:conclude} concludes the paper.

\section{Preliminaries}\label{SEC:preliminary}

Section \ref{SUBSEC:algoDiVinE} is adapted from \cite{VBBBipdps09divine}; the syntax outlined in Section \ref{SUBSEC:preDVE}, while incomplete, is meant for the presentation of the time-explicit description methods; the complete description can be found in \cite{divinePrjPage}.

\subsection{Distributed Model Checking Algorithms in {\sc DiVinE}}\label{SUBSEC:algoDiVinE}
{\sc DiVinE} is an explicit-state LTL model checker based on the automata-based procedure by Vardi and Wolper \cite{DBLP:conf/lics/VardiW86}. The property to be specified is described by an LTL formula. In LTL model checking, all efficient \emph{sequential} algorithms are based on the \emph{postorder} exploration as computed by a depth-first search (DFS) of the state space. However, computing DFS postorder is P-complete \cite{DBLP:journals/ipl/Reif85DFSisPcomplete}, so no benefit in terms of either time or space will result from parallelization of this type of algorithm.

Two algorithms, OWCTY and MAP \cite{Barnat05DivineAlgo}, are introduced in {\sc DiVinE}. The sequential complexity of each is worse than that of the DFS-based algorithms but both can be efficiently implemented in parallel. OWCTY, or \emph{One Way to Catch Them Young}, is based on the fact that a directed graph can be topologically sorted if and only if it is acyclic. The algorithm applies a standard linear topological sort algorithm to the graph. Failure in the sorting means the graph contains a cycle. Accepting cycles are detected with multiple rounds of the sorting. MAP, or \emph{Maximal Accepting Predecessors}, is based on the fact that each accepting vertex in an accepting cycle is its own predecessor. To improve memory efficiency, the algorithm only stores a single representative accepting predecessor for each vertex by choosing the maximal one in a linear ordering of vertices.

These two algorithms are preferable in different cases. If the property of a model is expected to hold, and the state space can fit completely into (distributed) memory, OWCTY is preferable as it is three times faster than MAP to explore the whole state space. On the other hand, MAP can generally find a counterexample (if it exists) more quickly as it works on-the-fly.

\subsection{{\sc DiVinE} Modeling Language}\label{SUBSEC:preDVE}
DVE is the modeling language of {\sc DiVinE}. Like in Promela (the modeling language of SPIN), a model described in DVE consists of processes, message channels and variables. Each process, identified by a unique name $procid$, consists of lists of local variable declarations and state declarations, the initial state declaration and a list of transitions.

A transition transfers the process state from $ stateid_{1}$ to $ stateid_{2}$. The transition may contain a guard (which decides whether the transition can be executed), a synchronization (which communicates data with another process) and an effect (which assigns new values to local or global variables). So we have

\

{\tt \ Transition ::= $ stateid_{1}$  -> $ stateid_{2}$  \{ Guard Sync Effect \} }

\

The {\tt Guard} contains the keyword {\tt guard} followed by a boolean expression and the {\tt Effect} contains the keyword {\tt effect} followed by a list of assignments. The {\tt Sync} follows the denotation for communication in CSP, `!' for the sender and `?' for the receiver. The synchronization can be either asynchronous or rendezvous. Value(s) is transferred in the channel identified by $chanid$. So we have

\

{\tt \ Sync ::= sync $chanid$ ! SyncValue $\vert$ $chanid$ ? SyncValue ;}

\

A \emph{property process} is automatically generated for the corresponding property written as an LTL formula. Modeled system processes and the property process progress synchronously, so the latter can observe the system's behavior step by step and catch errors.

\section{Explicit-Time Description Methods}\label{SEC:timeDesc}

With explicit-time description methods, the passage of time and timed quantified values can be expressed in un-timed languages and properties to be specified can be expressed in conventional temporal logics. This section describes Lamport's LEDM before detailing our new EEDM. At the end of this section, we study a small pre-emptive example with respect to explicit-time description methods.

\subsection{The Lamport Explicit-time Description Method}\label{SUBSEC:timeLamport}

In LEDM, current time is represented with a global variable \emph{now} that is incremented by an added \emph{Tick} process. As we mentioned earlier, standard model checkers can only deal with integer variables, and a real-time system can only be modeled in discrete-time using an explicit-time description. So the \emph{Tick} process increments \emph{now} by 1. Note that in explicit-time description methods for standard model checkers, the real-valued time variables must be replaced by integer-valued ones. Therefore, these methods in general do not preserve the continuous-time semantics; otherwise an inherently infinite-state specification will be produced and the verification will be undecidable. However, they are sound for a commonly used class of real-time systems and their properties \cite{DBLP:conf/icalp/HenzingerMP92}.

\begin{figure}[h!]
\begin{center}
  \includegraphics[width=3in]{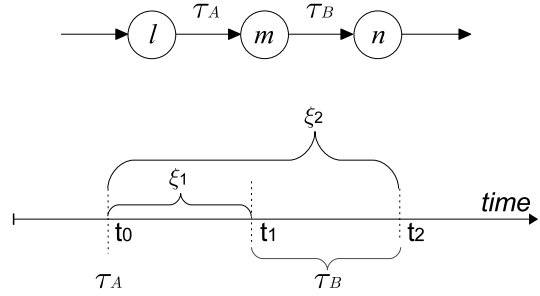}\\
  \caption{States and Timeline of process $P_{i}$}\label{Fig:timelinePi}
\end{center}
\end{figure}

Placing lower-bound and upper-bound timing constraints on transitions in processes is the common way to model real-time systems. Figure \ref{Fig:timelinePi} shows a simple example of only two transitions: transition $\tau_{A}$: {\tt $ stateid_{l}$  -> $ stateid_{m}$} is followed by the transition $\tau_{B}$: {\tt $ stateid_{m}$  -> $ stateid_{n}$}. An upper-bound timing constraint on when transition $\tau_{B}$ must occur is expressed by a guard on the transition in the \emph{Tick} process so as to prevent an increase in time from violating the constraint. A lower-bound constraint on when transition $\tau_{B}$ may occur is expressed by a guard on $\tau_{B}$ so it cannot be executed earlier than it should be. Each system process $P_{i}$ has a pair of count-down timers denoted as global variables $ubtimer_{i}$ and $lbtimer_{i}$ for the timing constraints on its transitions. A large enough integer constant, denoted as {\tt INFINITY}, is defined. All upper bound timers are initialized to {\tt INFINITY} and all lower bound timers are initialized to zero. Upper bound timers with the value of {\tt INFINITY} are not active and the \emph{Tick} process will not decrement them.  For transition $\tau_{B}$, the timers will be set to the correct values by $\tau_{A}$: {\tt $ stateid_{l}$  -> $ stateid_{m}$}. As \emph{now} is incremented by 1, each non{\tt -INFINITY} {\tt ubtimer} and non-zero {\tt lbtimer} is decremented by 1.

\begin{figure}[h!]

{\tt \hspace{16 pt} process P\_Tick \{ }

{\tt \hspace{30 pt} state tick; }

{\tt \hspace{30 pt} init tick; }

{\tt \hspace{30 pt} trans }

{\tt \hspace{50 pt} tick -> tick \{ guard $all$  $ubtimers$ > 0;  }

{\tt \hspace{128 pt} effect now = now + 1, }

{\tt \hspace{158 pt} $decrements$ $all$ $timers$; \} ; }

{\tt \hspace{16 pt} \}  }
\caption[]{\emph{Tick} process in DVE for LEDM \label{Fig:tickProcLamportMethod}}
\end{figure}

In Figure \ref{Fig:timelinePi}, initially, $(ubtimer_{i},lbtimer_{i})$ is set to $({\tt INFINITY},0)$. Transition $\tau_{A}$ is executed at time instant $t_{0}$, and $(ubtimer_{i},lbtimer_{i})$ is set to $(\xi_{2},\xi_{1})$. After $\xi_{1}$ time units, i.e., at time instant $t_{1}$ when $(ubtimer_{i},lbtimer_{i})$ is equal to $(\xi_{2}-\xi_{1},0)$, transition $\tau_{B}$ is enabled. Both timers will be reset or set to new time bounds after the execution of $\tau_{B}$. If transition $\tau_{B}$ is still not executed when the time reaches $t_{2}$ and $ubtimer_{i}$ is equal to 0, the transition in the \emph{Tick} process is disabled. This forces transition $\tau_{B}$ (it is the only transition possible at this time) to set the $ubtimer_{i}$; then the \emph{Tick} process can start again. In this way, the time upper-bound constraint is realized. The \emph{Tick} process and the system process $P_{i}$ in DVE are described in Figure \ref{Fig:tickProcLamportMethod} and Figure \ref{Fig:sysProcLamportMethod}.

\begin{figure}[h!]
{\tt \hspace{16 pt} process P\_i \{ }

{\tt \hspace{30 pt} state ..., state\_l, state\_m, state\_n; }

{\tt \hspace{30 pt} init ...; }

{\tt \hspace{30 pt} trans }

{\tt \hspace{70 pt} ... -> ... , }

{\tt \hspace{50 pt} state\_l -> state\_m \{ ...; }

{\tt \hspace{150 pt} effect $set$ $timers$ $for$ $transition\ \tau_{B}$;\}, }

{\tt \hspace{50 pt} state\_m -> state\_n \{ guard lbtimer$[i]$==0 ; effect ... ; \}, }

{\tt \hspace{70 pt} ... -> ... ; }

{\tt \hspace{16 pt} \}  }
\caption[]{System process $P_{i}$ in DVE for LEDM \label{Fig:sysProcLamportMethod}}
\end{figure}

We observe that the value of \emph{now} is limited by the size of type {\tt integer} and careless incrementing can cause overflow error. This can be avoided by incrementing \emph{now} using modular arithmetic, i.e., setting $now = (now+1)$ {\tt mod MAXIMAL} ({\tt MAXIMAL} is the maximal integer value supported by the model checker). The value limit can also be increased by linking several integers, i.e., every time {\tt ($int_1$+1) mod MAXIMAL} becomes zero again,  $int_2$ increments by 1, and so on. Note that the variable \emph{now} is only incremented in the \emph{Tick} process and does not appear in any other process. So for general system models in which time lower and upper bounds suffice, the variable \emph{now} should be removed.

\subsection{The New Efficient Explicit-Time Description Method}\label{SUBSEC:timeEEDM}

This section is organized as follows. First, we describe the leaping mode and the standard mode of the new EEDM in section \ref{SUBSUBSEC:LeapingTickEEDM} and \ref{SUBSUBSEC:StandardTickEEDM} respectively. Second, we present some discussions (clarifications) of issues on EDMs and EEDM in section \ref{SUBSUBSEC:EEDMissues}. Finally, a pre-emptive scheduling modeling example using EEDM is described in section \ref{SUBSUBSEC:withTimedAutomata}.

\subsubsection{Leaping Ticks}\label{SUBSUBSEC:LeapingTickEEDM}

\begin{figure}[h!]
\begin{center}
  \includegraphics[width=3in]{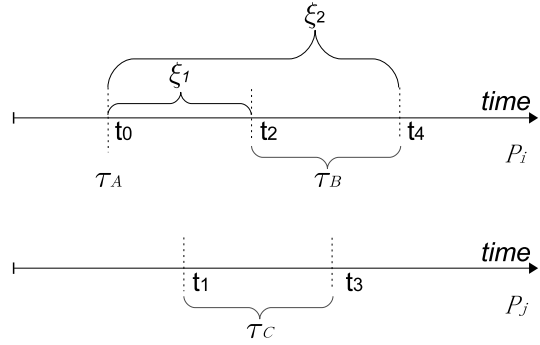}\\
  \caption{Timeline of process $P_{i}$ and $P_{j}$}\label{Fig:timelinePiPj}
\end{center}
\end{figure}

All aforementioned explicit-time description methods (LEDM, SEDM and SMEDM) increase $now$ by 1 each tick. On the other hand, consider Figure \ref{Fig:timelinePiPj}: we observe that when the system contains \emph{only} one process, $P_{i}$, after $t_{0}$, $\tau_{B}$ cannot be executed until time reaches $t_{2}$. Therefore, the ticks between $t_{0}$ and $t_{1}$ serve no purpose; optimally, the \emph{Tick} process should directly ``leap'' to $t_{2}$. Similarly, $\tau_{B}$ is enabled between $t_{2}$ and $t_{4}$, so either $\tau_{B}$ is executed before $t_{4}$ or time reaches $t_{4}$ and $\tau_{B}$'s execution is forced; therefore, the \emph{Tick} process can leap to $t_{4}$ from $t_{2}$. When we include $P_{j}$, after $t_{0}$, the \emph{Tick} should first leap to $t_{1}$ so $P_{j}$ can enable transition $\tau_{C}$; then it should leap to $t_{2}$ and so on.

Based on these observations, in the new EEDM, we use one global count-down timer for each system process, e.g., $timer_{i}$ for $P_{i}$ in Figure \ref{Fig:timelinePiPj} is set to $\xi_{1}$ on $t_{0}$ and to $\xi_{2}-\xi_{1}$ on $t_{2}$. The \emph{Tick} process increments \emph{now} by the value of the smallest timer on condition that no timer equals zero and at least one timer is non{\tt -INFINITY}. In fact, the \emph{Tick} process, leaping in this way, is running in the \emph{leaping} mode; the \emph{Tick} process in leaping mode and the corresponding system process $P_{i}$ in DVE are described in Figure \ref{Fig:tickProcEEDM1} and Figure \ref{Fig:systemProcEEDM1} ($N$ is the number of system processes).

\begin{figure}[h!]
{\tt \hspace{16 pt} process P\_Tick \{ }

{\tt \hspace{30 pt} state tick; }

{\tt \hspace{30 pt} init tick; }

{\tt \hspace{30 pt} trans }

{\tt \hspace{50 pt} tick -> tick \{ guard  $(\wedge_{1..N}({\tt timer}[i]>0))\wedge(\vee_{1..N}({\tt timer}[i]\neq{\tt INFINITY}))$; }

{\tt \hspace{125 pt} effect $now=now+\min_{1..N}({\tt timer}[i])$, }

{\tt \hspace{160 pt} $decrement$ $all$ $timers$ $by$ $\min_{1..N}({\tt timer}[i])$;\}; }

{\tt \hspace{16 pt} \}  }
\caption[]{\emph{Tick} process in leaping mode in DVE for EEDM \label{Fig:tickProcEEDM1}}
\end{figure}

\begin{figure}[h!]
{\tt \hspace{16 pt} process P\_i \{ }

{\tt \hspace{30 pt} state state\_l, state\_m1, state\_m2, state\_n, ...; }

{\tt \hspace{30 pt} init ...; }

{\tt \hspace{30 pt} trans }

{\tt \hspace{70 pt} ... \  -> \ ... , }

{\tt \hspace{50 pt} state\_l \ -> state\_m1 \{ ...; effect {\tt timer}[i]=$\xi_{1}$;\},}

{\tt \hspace{50 pt} state\_m1 -> state\_m2 \{ guard {\tt timer}[i]=0; effect {\tt timer}[i]=$\xi_{2}-\xi_{1}$; \}, }

{\tt \hspace{50 pt} state\_m2 -> state\_n \{ $executes\ \tau_{B}$ $and$ $resets\ {\tt timer}[i]$; \}, }

{\tt \hspace{70 pt} ... \  -> \ ... ; }

{\tt \hspace{16 pt} \}  }
\caption[]{System process $P_{i}$ in DVE for EEDM \label{Fig:systemProcEEDM1}}
\end{figure}

\subsubsection{To Know the Current Time Instant}\label{SUBSUBSEC:StandardTickEEDM}

Careful readers may notice that there is one penalty for \emph{Tick} to leap: the actual time instant when $\tau_{B}$ is executed is unknown unless it is at $t_{4}$. In fact, in the leaping mode, it is only known that a transition is executed between the two closest ticks that nest the transition. Consider the example in Figure \ref{Fig:timelinePiPj}; the \emph{Tick} will sequentially leap from $t_{0}$ through $t_{4}$; $\tau_{B}$ may be executed on: (1) some time instant between $t_{2}$ and $t_{3}$; or (2) some time instant between $t_{3}$ and $t_{4}$; or (3) the time instant of $t_{4}$. However, as we discussed earlier in Section \ref{SEC:introduction} and in \cite{Hao09SMEDM}, in many systems, especially those with pre-emptive scheduling, it is necessary to know the actual time instant when the transition is executed.

\begin{figure}[h!]
{\tt \hspace{8 pt} process P\_Tick \{ }

{\tt \hspace{21 pt} state tick; }

{\tt \hspace{21 pt} init tick; }

{\tt \hspace{21 pt} trans }

{\tt \hspace{35 pt} tick -> tick \{ }

{\tt \hspace{46 pt} guard  $(\wedge_{1..N}({\tt timer}[i]>0))\wedge(\vee_{1..N}({\tt timer}[i]\neq{\tt INFINITY}))\wedge(\wedge_{1..N}({\tt signal}[i]==0))$; }

{\tt \hspace{46 pt} effect $now=now+\min_{1..N}({\tt timer}[i])$, }

{\tt \hspace{81 pt} $decrement$ $all$ $timers$ $by$ $\min_{1..N}({\tt timer}[i])$;\}, }

{\tt \hspace{35 pt} tick -> tick \{ }

{\tt \hspace{46 pt} guard  $(\wedge_{1..N}({\tt timer}[i]>0))\wedge(\vee_{1..N}({\tt timer}[i]\neq{\tt INFINITY}))\wedge(\vee_{1..N}({\tt signal}[i]==1))$; }

{\tt \hspace{46 pt} effect $now=now+1$, }

{\tt \hspace{81 pt} $decrement$ $all$ $timers$ $by$ $1$;\}; }

{\tt \hspace{8 pt} \}  }
\caption[]{\emph{Tick} process in standard mode in DVE for EEDM \label{Fig:tickProcEEDM2}}
\end{figure}

\begin{figure}[h!]
{\tt \hspace{10 pt} process P\_i \{ }

{\tt \hspace{24 pt} state state\_l, state\_m1, state\_m2, state\_n, ...; }

{\tt \hspace{24 pt} init ...; }

{\tt \hspace{24 pt} trans }

{\tt \hspace{64 pt} ... \  -> \ ... , }

{\tt \hspace{44 pt} state\_l \ -> state\_m1 \{ ...; effect {\tt timer}[i]=$\xi_{1}$;\},}

{\tt \hspace{44 pt} state\_m1 -> state\_m2 \{ guard {\tt timer}[i]=0; }

{\tt \hspace{160 pt} effect {\tt timer}[i]=$\xi_{2}-\xi_{1}$, {\tt signal}[i]=1; \}, }

{\tt \hspace{44 pt} state\_m2 -> state\_n \{ $executes\ \tau_{B}$ $and$ $resets\ {\tt timer}[i]$, {\tt signal}[i]=0; \}, }

{\tt \hspace{64 pt} ... \  -> \ ... ; }

{\tt \hspace{10 pt} \}  }
\caption[]{System process $P_{i}$ to illustrate the standard mode \label{Fig:systemProcEEDM2}}
\end{figure}

To overcome this problem, we allow the \emph{Tick} process to run in the \emph{standard} mode. We define a global signal variable for each system process. All signals are set to 0 at the initial state. Whenever a system process $P_{i}$ requires the current time for future calculation, $signal_{i}$ should be set to 1; the \emph{Tick} process in turn will run in the standard mode with which it will increment \emph{now} by 1 in each tick. E.g., when time reaches $t_{2}$ in Figure \ref{Fig:timelinePiPj}, $P_{i}$'s signal $signal_{i}$ is set to 1 in order to store the time instant at which $\tau_{B}$ is executed; when time reaches $t_{4}$, $signal_{i}$ is set back to 0 so that the \emph{Tick} switches back to leaping mode. Both the \emph{Tick} process and the system process need to be updated to incorporate the standard mode, see Figure \ref{Fig:tickProcEEDM2} and Figure \ref{Fig:systemProcEEDM2}.


\subsubsection{Issues on EDMs and EEDM}\label{SUBSUBSEC:EEDMissues}

Readers may be concerned about the verification capability of explicit-time description methods. As in our earlier discussion, EDMs simulate a \emph{discrete} timer by making use of existing constructs in standard un-timed model checkers; in other words, time is just another normal variable in an un-timed model. Therefore, EDMs are not affected by verification issues such as whether the property is specified as an LTL or CTL formula or whether the property is verified using explicit-state based (e.g., Spin) or symbolic model checking (e.g., SMV) algorithms. These verification issues depend on what standard un-timed model checker is used.

Discrete timed model checkers suffer from a common problem: how to find the right time quantum (granularity) that does not mask errors. E.g., for processes in a hospital, a time unit defined as a day will definitely mask an error which violates the property ``the patient receives a certain treatment within 1 hour''. On the other hand, the state space can easily blow up if a finer time unit is used. Readers may be concerned that the introduction of leaping ticks may add to this problem. Actually, leaping ticks do not musk errors in this aspect. The difference between LEDM and EEDM in leaping mode is that EEDM in leaping mode cannot record and use the exact time instant when a transition is executed in the model or the specified properties. For example, the LTL property that $b$ becomes true before 10 time units have elapsed since $\tau_{B}$ is executed cannot be verified using EEDM in leaping mode. For this reason, we introduce the mode-switching mechanism in EEDM.

To reduce the state space, Lamport \cite{Lamport05TRrealSimple} proposed the use of view symmetry, which is equivalent to abstraction for a symmetric specification \emph{S}. Abstraction consists of checking \emph{S} by model checking a different specification \emph{A} called an abstraction of \emph{S}. This technique has two restrictions: (1) the \emph{now} variable must be eliminated, which means the current time instant is not accessible in this case; (2) if the model checker does not support checking under view symmetry or abstraction, the abstraction specification \emph{A} must be constructed by hand. In addition, this reduction technique is orthogonal to our EEDM, i.e., we can use Lamport's abstraction technique in conjunction with EEDM.

The idea of leaping ticks in EEDM is quite similar to the notion of time regions in time-automata-based model checkers, which advances time up to the point where a transition must be executed in order not to violate the invariant defined on the corresponding state. However, the implementations are fundamentally different: time-automata-based model checkers introduce specialized data structures \cite{DBLP:conf/tacas/KrcalY04decidableTA} to store time regions and use symbolic model checking algorithms extended for time; on the other hand, EEDM, as with LEDM, only uses an explicit \emph{tick} process and some global variables, and the leaping way of advancing time is obtained by letting the tick leap to the next closest time bound of all systems processes.

\subsubsection{To Know The Current Time Instant: A Pre-emptive Scheduling Example}\label{SUBSUBSEC:withTimedAutomata}

Following the triage example described in Section \ref{SEC:introduction}, we consider a system of multiple parallel tasks with different priorities, assuming that the right to an exclusive resource is deprivable, i.e., a higher priority task \emph{B} may deprive the resource from the currently running task \emph{A}. In this case, the elapsed time of \emph{A}'s execution must be stored for a future resumed execution.
\begin{figure}[h!]
\begin{center}
  \includegraphics[width=3in]{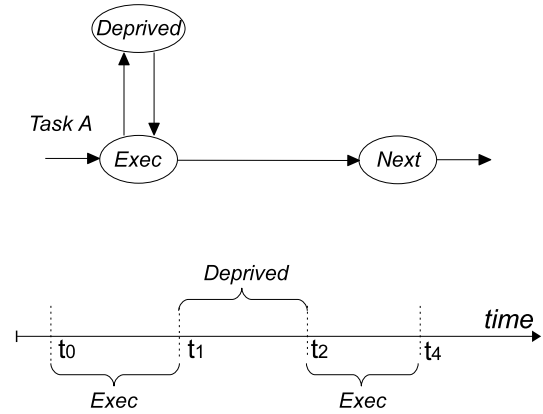}\\
  \caption{An Example Case of Pre-emptive Scheduling}\label{Fig:dynamicScheduling}
\end{center}
\end{figure}

\begin{figure}[h!]
{\tt \hspace{16 pt} byte isROccupied=0; $//$0 means available}

{\tt \hspace{16 pt} process A \{ }

{\tt \hspace{30 pt} default(Tag,$tag_{A}$) }

{\tt \hspace{30 pt} int timeToGo=10; }

{\tt \hspace{30 pt} state s\_i, s\_Exec, s\_Deprived, ...; }

{\tt \hspace{30 pt} init ...; }

{\tt \hspace{30 pt} trans }

{\tt \hspace{50 pt} ...  -> ... ; }

{\tt \hspace{50 pt} s\_i -> s\_Exec \{  }

{\tt \hspace{120 pt} guard isROccupied==0;  }

{\tt \hspace{120 pt} effect isROccupied=Tag, timer[A]=timeToGo, signal[A]=1; }

{\tt \hspace{50 pt} s\_Exec -> s\_Deprived \{ }

{\tt \hspace{120 pt} guard isROccupied\!=Tag \&\& timer[A]>0; }

{\tt \hspace{120 pt} effect timeToGO=timer[A]; \}, }

{\tt \hspace{50 pt} s\_Deprived -> s\_Exec \{ }

{\tt \hspace{120 pt} guard isROccupied==0; }

{\tt \hspace{120 pt} effect isROccupied=Tag, timer[A]=timeToGo; \}, }

{\tt \hspace{50 pt} s\_Exec -> s\_Next \{ }

{\tt \hspace{120 pt} guard timer[A]==0; }

{\tt \hspace{120 pt} effect isROccupied=0, signal[A]=0; \}, }

{\tt \hspace{50 pt} ...  -> ... ; }

{\tt \hspace{16 pt} \}  }
\caption[]{Process in DVE for Pre-emptive Scheduling Example using EEDM} \label{Fig:ProcDynamic}
\end{figure}

Figure \ref{Fig:dynamicScheduling} shows a portion of a state transition diagram for task \emph{A}, assuming \emph{A} needs the exclusive resource \emph{R} for 10 time units; when \emph{R} becomes available at time instant $t_{0}$, \emph{A} starts its execution by entering the state \emph{Exec}; at time instant $t_{1}$, \emph{B} deprives \emph{A}'s right to \emph{R}, and \emph{A} changes to the state \emph{Deprived} and stores the elapsed $t_{1}-t_{0}$ time units; when \emph{R} becomes available again, \emph{A} resumes it execution to state \emph{Exec} for the remaining $10-(t_{1}-t_{0})$ units. Implementation of this example using any one of the three explicit-time description methods is straightforward. Figure \ref{Fig:ProcDynamic} shows the process for task \emph{A} in DVE using EEDM (assuming \emph{A} has the lowest priority).

\section{Experiments}\label{SEC:exprDesc}

\subsection{Overview}\label{SUBSEC:fischerAlgo}

For the convenience of comparison with LEDM in DiVinE, we use the Fischer's mutual exclusion algorithm as in \cite{Hao09SEDM} \cite{Hao09SMEDM}; this algorithm is a well-known benchmark for timed model checking. The description of the algorithm below is adapted from \cite{Lamport05TRrealSimple}. Our experiment is to model the algorithm in {\sc DiVinE} using EEDM in both standard and leaping modes, and compare the time and memory efficiency and size of state space with that of LEDM (we omit the experiments for SEDM and SMEDM because they are comparable with LEDM in the aforementioned three numeric criteria).

Fischer's algorithm is a shared-memory, multi-threaded algorithm. It uses a shared variable \emph{x} whose value is either a thread identifier (starting from 1) or zero; its initial value is zero. For the convenience of specification of the safety property in our experiments, we use a counter \emph{c} to count the number of threads that are in the critical section. The program for thread \emph{t} is described in Figure \ref{Fig:fischerAlgo}.

\begin{figure}[h!]
{\parindent170pt  {\it ncs}: noncritical section;

  {\it \ \ \ a}: {\bf wait until} {\it x} = 0;

  {\it \ \ \ b}: {\it x} := {\it t};

  {\it \ \ \ c}: {\bf if} {\it x} $\neq$ {\it t} {\bf then goto} {\it a};

  {\it \ \ cs}: critical section;

  {\it \ \ \ d}: {\it x} := 0; {\bf goto} {\it ncs};
}
\caption[]{Program of thread \emph{t} in Fischer's algorithm \label{Fig:fischerAlgo}}
\end{figure}

The timing constraints are: first, step \emph{b} must be executed at most $\delta_{b}^{u}$ time units (as an upper bound) after the preceding execution of step \emph{a}; second, step \emph{c} cannot be executed until at least $\delta_{c}^{l}$ time units (as a lower bound) after the preceding execution of step \emph{b}. For step \emph{c}, there is an additional upper bound $\delta_{c}^{u}$ to ensure fairness, i.e., step \emph{c} will eventually be executed. The algorithm is tested for 6 threads. The safety property to be verified, {\it ``no more than one process can be in the critical section''}, is specified as $G (c<2)$ for the model.

Version 0.8.1 of the {\sc DiVinE}-Cluster is used. This version has the new feature of pre-compiling the model in DVE into dynamically linked C functions; this feature speeds up the state space generation significantly. As the example property is known to hold, the OWCTY algorithm is chosen for better time efficiency.

All experiments are executed on the Mahone cluster of ACEnet \cite{ACEnetPrjPage}, the high performance computing consortium for universities in Atlantic Canada. The cluster is a Parallel Sun x4100 AMD Opteron (dual-core) cluster equipped with Myri-10G interconnection. Parallel jobs are assigned using the Open MPI library.

\subsection{Experiment 1}\label{SUBSEC:expr1}

For the first experiment, we use the same value for three constraints, i.e., $\delta_{b}^{u}=\delta_{c}^{l}=\delta_{c}^{u}=T$. Figure \ref{Fig:timeResults1} compares time and memory efficiency for the two explicit-time description methods with 16 CPUs.

\begin{figure*}[h!]
\begin{center}
\begin{tabular}{|r|r|r|r|r|r|r|r|r|r|}
  \hline
      & \multicolumn{3}{c|}{LEDM} & \multicolumn{6}{c|}{EEDM} \\ \cline{5-10}
      & \multicolumn{3}{c|}{\ }    & \multicolumn{3}{c|}{standard} & \multicolumn{3}{c|}{leaping} \\ \cline{2-10}
  \emph{T} & {States} & {Time}  & {Memory} & {States} & {Time}  & {Memory} & {States} & {Time} & {Memory} \\ \hline
  2 & 644,987 & 1.8 & 4,700.1 & 626,312 & 1.9 & 4,689.6 & 141,695 & 1.4 & 4,606.2 \\
  3 & 1,438,204 & 2.4 & 4,822.3 & 2,375,451 & 3.4 & 4,982.7 & 141,695 & 1.5 & 4,612.8 \\
  4 & 3,048,515 & 3.3 & 4,942.8 & 7,363,766 & 5.0 & 5,820.9 & 141,695 & 1.5 & 4,603.6 \\
  5 & 6,033,980 & 4.2 & 5,603.4 & 19,471,191 & 10.4 & 7,855.2 & 141,695 & 1.4 & 4,604.9 \\
  6 & 11,201,179 & 7.2 & 6,343.4 & 45,552,076 & 24.4 & 12,241.1 & 141,695 & 1.4 & 4,620.6 \\
  7 & 19,671,092 & 11.1 & 7,885.7 & 96,871,373 & 52.1 & 20,663.7 & 141,695 & 1.6 & 4,605.7 \\
  8 & 32,952,899 & 18.6 & 9,958.9 & 190,941,594 & 133.0 & 37,503.6 & 141,695 & 1.4 & 4,601.8 \\
  9 & 53,025,700 & 30.2 & 13,288.7 & 353,811,115 & 246.5 & 63,572.8 & 141,695 & 1.4 & 4,622.4 \\ \hline
\end{tabular}
\caption[]{Number of states, Time (in seconds) and memory usage (in MB) for Experiment 1 \label{Fig:timeResults1}}
\end{center}
\end{figure*}

We can see the significant advantage of EEDM in leaping mode: the number of states, verification time and memory usage remain virtually the same for all \emph{T}s. Remark that all timing bounds are the same for all threads; the \emph{Tick} process always leaps $T$ time units in each tick (it ticks only when there is at least one active timer). Therefore, changing the value of $T$ will not change the number of states. 

Now we compare LEDM and EEDM in standard mode. Let ${\tt states}(X)$ be the number of states of method $X$. We can see that, after $T=3$, ${\tt states(EEDM_{standard})}>{\tt states(LEDM)}$. As $T$ increases from 2 to 9, ${\tt states(EEDM_{standard})}$ increases by a factor of 564.9 while ${\tt states(LEDM)}$ increases by a factor of only 82.2; a comparison of the verification time yields similar results. The system process in EEDM has more transitions than LEDM because there is only one timer for each system process and a timer needs to be assigned twice if the next transition has both lower and upper bounds (e.g. $\tau_{B}$ of $P_{i}$ in Figure \ref{Fig:timelinePiPj}, {\tt timer[i]} is assigned to be $\xi_{1}$ and $\xi_{2}-\xi_{1}$ at $t_{0}$ and $t_{2}$ respectively); on the other hand, LEDM has two timers for each system process so assigning both bounds can be made in one step.

\subsection{Experiment 2}\label{SUBSEC:expr2}

\begin{figure}[h!]
\begin{center}
  \includegraphics[width=6in]{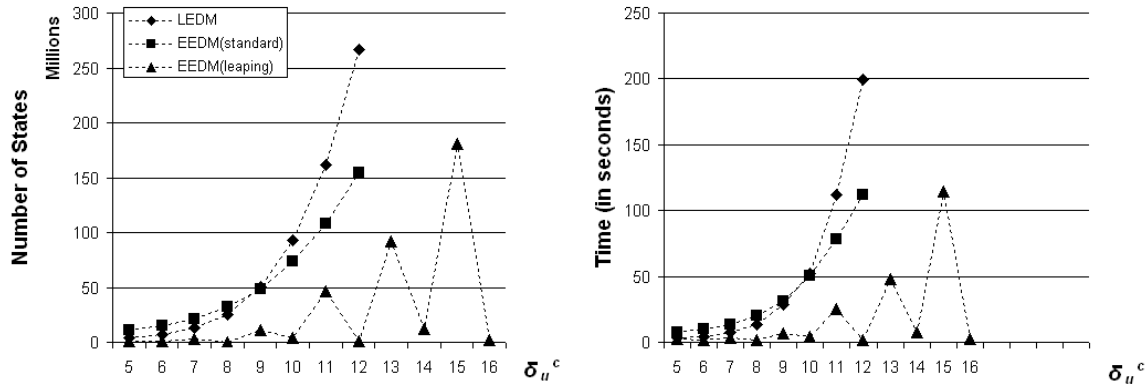}\\
  \caption{Number of states and Time (in seconds) for Experiment 2}\label{Fig:timeResultsChart2}
\end{center}
\end{figure}

For the second experiment, we set $\delta_{b}^{u}$ and $\delta_{c}^{l}$ to 4 and vary $\delta_{c}^{u}$. Figure \ref{Fig:timeResults2} compares the number of states, time and memory efficiency for the two explicit-time description methods with 16 CPUs. Figure \ref{Fig:timeResultsChart2} shows how the size of the state space and verification time grow as $\delta_{c}^{u}$ increases. The extra experimental data for $\delta_{c}^{u}=\{13,14,15,16\}$ are intended to articulate the growing pattern of the state space of EEDM in leaping mode.

\begin{figure*}[h!]
\begin{center}
\begin{tabular}{|r|r|r|r|r|r|r|r|r|r|}
  \hline
      & \multicolumn{3}{c|}{LEDM} & \multicolumn{6}{c|}{EEDM} \\ \cline{5-10}
      & \multicolumn{3}{c|}{\ }    & \multicolumn{3}{c|}{Standard} & \multicolumn{3}{c|}{Leaping} \\ \cline{2-10}
  $\delta_{c}^{u}$ & {States} & {Time}  & {Memory} & {States} & {Time}  & {Memory} & {States} & {Time} & {Memory} \\ \hline
  5 & 3,659,317 & 3.5 & 5,199.1 & 10,865,877 & 7.2 & 6,415.6 & 1,122,491 & 2.2 & 4,771.0 \\
  6 & 6,783,455 & 4.2 & 5,770.2 & 15,221,140 & 10.2 & 7,150.3 & 1,046,759 & 2.0 & 4,758.0 \\
  7 & 12,907,369 & 7.2 & 6,754.2 & 21,451,024 & 13.2 & 8,198.5 & 3,516,193 & 3.6 & 5,182.7 \\
  8 & 25,723,697 & 13.3 & 8,898.8 & 31,934,332 & 20.2 & 9,946.8 & 365,279 & 1.6 & 4,651.1 \\
  9 & 50,500,739 & 28.2 & 13,047.6 & 48,889,270 & 31.2 & 12,721.1 & 10,998,335 & 7.1 & 6,434.9 \\
  10 & 93,349,553 & 52.3 & 20,146.1 & 73,501,090 & 50.7 & 16,858.4 & 3,828,687 & 3.8 & 5,228.0 \\
  11 & 161,886,059 & 111.9 & 31,722.6 & 108,005,926 & 78.5 & 23,104.9 & 46,149,106 & 24.9 & 12,313.8 \\
  12 & 266,256,377 & 199.2 & 49,154.8 & 154,662,946 & 112.2 & 30,045.6 & 857,773 & 1.9 & 4,735.3 \\
  13 &  &  &  &  &  &  & 92,147,198 & 48.4 & 19,928.2 \\
  14 &  &  &  &  &  &  & 12,275,835 & 7.3 & 6,650.4 \\
  15 &  &  &  &  &  &  & 180,459,742 & 114.1 & 34,098.7 \\
  16 &  &  &  &  &  &  & 1,847,395 & 2.7 & 4,911.5 \\
  \hline
\end{tabular}
\caption[]{Number of states, Time (in seconds) and memory usage (in MB) for Experiment 2 \label{Fig:timeResults2}}
\end{center}
\end{figure*}

As opposed to the results in experiment 1, in this experiment EEDM in standard mode performs better than LEDM. We can see that after $\delta_{c}^{u}=9$, ${\tt states(EEDM_{standard})}<{\tt states(LEDM)}$; as the model becomes larger, ${\tt states(EEDM_{standard})}$ increases more slowly than ${\tt states(LEDM)}$. In fact, as $\delta_{c}^{u}$ increases from 5 to 12, ${\tt states(LEDM)}$ increases by a factor of 72.8 while ${\tt states(EEDM_{standard})}$ increases by a factor of only 14.2; we can see similar comparison results in terms of the verification time.

EEDM in leaping mode still shows much better performance than LEDM and EEDM in standard mode; ${\tt states(EEDM_{leaping})}$ also shows an interesting phenomenon as $\delta_{c}^{u}$ increases. The number of states of both EEDM in standard mode and LEDM increase at a relatively more steady speed: as $\delta_{c}^{u}$ increases by 1, ${\tt states(EEDM_{standard})}$ increases by a factor of about 1.45 and ${\tt states(LEDM)}$ increases by a factor of about 1.8. On the other hand, the increments of ${\tt states(EEDM_{leaping})}$ are grouped by the value of $s=(\delta_{c}^{u}\ {\tt mod}\ \delta_{c}^{l})$. We can see that, for the same $\lfloor {\delta_{c}^{u} \over \delta_{c}^{l}} \rfloor$, ${\tt states(EEDM_{leaping})_{s=0}}<{\tt states(EEDM_{leaping})_{s=2}}<{\tt states(EEDM_{leaping})_{s=1}}<{\tt states(EEDM_{leaping})_{s=3}}$. For $s=0$, whenever there is more than one active timer, their values are integer multiples of $\delta_{c}^{l}$ (4 in this experiment), so the \emph{Tick} still leaps at least 4 time units each tick; in the case of $s=2$, the \emph{Tick} leaps at least 2 time units each tick. On the other hand, for $s=1$ and $s=3$, in the worst case, the \emph{Tick} leaps only 1 time units each tick. From these observations, we can conclude that EEDM in leaping mode performs better the greater the \emph{greatest common divisor} (gcd) of all timing bounds of all system processes.
%

\section{Conclusion}\label{SEC:conclude}

In this paper, we present a new explicit-time description method, Efficient Explicit-time Description Method (EEDM) which is significantly more efficient than LEDM, SEDM and SMEDM. In addition to the improved efficiency, EEDM still retains the ability to store and access the current time for future calculations in the system model. Altogether, we have devised methods that have advantages in different aspects of real-time modeling: SEDM and SMEDM have better modularity and adaptability; EEDM is more efficient. These explicit-time description methods provide systematic ways to represent discrete time in un-timed model checkers like SPIN, SMV and {\sc DiVinE}.

In fact, the explicit-time description methods are intended to offer more options for the verification of real-time systems. First, as Van den Berg et al. mention in \cite{DBLP:conf/fmics/BergSW07LEDMcaseStudy}, in some real-world scenarios when significant resources have been invested into the model for a standard model checker, it is much easier and therefore preferable to extend the existing model to represent time notions rather than re-modeling the entire system for a specialized timed model checker. Second, explicit-time description methods provide a solution for accessing and storing the current clock value for timed-automata-based model checkers. Last and most important, explicit-time description methods, especially the EEDM, enable the usage of large-scale distributed model checkers so that we can verify much bigger real-time systems.

This research is part of an ambitious research and development project, Building Decision-support through Dynamic Workflow Systems for Health Care \cite{keith09CareMS}. Real world workflow processes can be highly dynamic and complex in a health care setting. Verification that the system meets its specifications is essential. Standard workflow patterns are widely used in business processes modeling, so we have translated most of the control-flow patterns into DVE and applied them in verifying two small process models \cite{Mashiyat09Pattern}. As a continuous effort, we will incorporate explicit-time description methods into workflow patterns' DVE specification and verify a larger model of the real-world healthcare processes with timing information.

As a more complex case study of EEDM, we are now building a pre-emptive scheduling model in the setting of the Dynamic Voltage Scaling (DVS) technique. We also plan to study the possibility of applying different abstraction techniques to the explicit-time description methods: Dutertre and Sorea \cite{DBLP:DutertreS04caldrAutomata} and Clarke et al. \cite{clarke07abstraction} recently presented two different abstraction techniques for timed automata and the abstraction outcome can be verified using un-timed model checkers.

\section*{Acknowledgment}
This research is sponsored by Natural Sciences and Engineering Research Council of Canada (NSERC), an Atlantic Computational Excellence Network (ACEnet) Post Doctoral Research Fellowship and the Atlantic Canada Opportunities Agency (ACOA) through an Atlantic Innovation Fund project. The computational facilities are provided by ACEnet. We thank Jiri Barnat, Keith Miller and the anonymous reviewers of PDMC'09 for their valuable comments.

\bibliographystyle{eptcs}
\bibliography{explicitTimeDesc3rd}

%
%
%
%
%
%
%
%
%
%
%
%
%
%
%
%
%
%
%
%
%
%

\end{document}